\documentclass{optica-article}

\journal{opticajournal} 

\articletype{Research Article}

\usepackage{lineno,braket}

\begin{document}

\title{Pure narrowband photon-pair generation in a monolithic cavity}

\author{Xavier Barcons Planas,\authormark{1,2,*} Helen M. Chrzanowski,\authormark{2,1} and Janik Wolters \authormark{2,3,4}}

\address{
\authormark{1}Institut für Physik, Humboldt-Universität zu Berlin, Newtonstr. 15, 12489 Berlin, Germany\\
\authormark{2}Institute of Space Research, German Aerospace Center, Rutherfordstr. 2, 12489 Berlin, Germany\\
\authormark{3}Institut für Physik und Astronomie, Technische Universität Berlin, Hardenbergstr. 36, 10623 Berlin, Germany\\
\authormark{4}AQLS UG (haftungsbeschränkt), Guerickestr. 12, 10587 Berlin, Germany}

\email{\authormark{*}xavier.barconsplanas@dlr.de} 


\begin{abstract*} 
Photonic quantum technologies require efficient sources of pure single photons. We present a heralded single-photon source based on spontaneous parametric down-conversion in a monolithic cavity optimized for high spectral and spatial purity. The source heralds single photons at a  wavelength of 1540~nm and a spectral bandwidth of 168~MHz, with a maximum heralding efficiency of 70\% including all transmission and detection losses,  while keeping the multi-photon contamination below 3\%. The cavity enhancement predominantly generates photons into the central cavity mode, with a theoretical upper bound on the spectral purity of 79.4\% arising from nonzero overlap with adjacent cavity modes. Spectral isolation of the central cavity mode with an etalon yields an increased measured spectral purity of $(96.2\pm2.7)\%$.
\end{abstract*}

\section{Introduction}
Second-generation quantum technologies leverage quantum information to achieve capabilities unattainable with classical systems~\cite{nielsen2010quantum}, including quantum computation, communication, and metrology. Among the candidates for carriers of quantum information, single photons are particularly well-suited~\cite{o2009photonic} due to their low decoherence and light-speed transmission. Many quantum information processing applications with single photons rely on the fundamental Hong-Ou-Mandel (HOM) two-photon interference~\cite{hong1987measurement,pan2012multiphoton,bouchard2020two}. This phenomenon is essential for applications, such as quantum networks with repeaters~\cite{azuma2023quantum}, photonic quantum computing~\cite{o2007optical}, or generation of heralded multiphoton entangled states via fusion operations~\cite{browne2005resource,li2021heralded,cao2024photonic}. Since the strength of the two-photon interference effect is ultimately limited by the purity of the photons involved, the generation of pure, indistinguishable single photons with high efficiency is a crucial requirement for advancing the performance and scalability of quantum information processing with light~\cite{wang2025scalable}. 

Outstanding among the most mature techniques for producing single photons is heralding them from photon-pair sources based on spontaneous parametric down-conversion (SPDC)~\cite{christ2013parametric}, owing to benefits including room-temperature operation, implementation simplicity, robustness, and replicability. Nevertheless, the heralded photon is not typically emitted into a pure state due to spatio-spectral correlations~\cite{walborn2010spatial,grice1997spectral} arising from energy and momentum conservation between the photons of the pair. As a result, the collection and subsequent detection of the herald photon in a specific spatio-temporal mode projects its partner photon into a mixed state of spatial and spectral modes, reducing both the spectral purity and heralding efficiency~\cite{meyer2017limits}.

Intensive efforts have been undertaken to generate photons from SPDC with pure-state characteristics. The simplest solution is to apply spatial and spectral filtering at the cost of high optical losses and reduced brightness. Alternatively, careful engineering of the SPDC process can restrict the emission without the use of filtering: waveguides can confine the emission of the down-converted photons to the fundamental spatial mode due to tight confinement of the fields~\cite{christ2009spatial}, while group velocity matching techniques~\cite{grice2001eliminating,uren2005generation,mosley2008heralded} or SPDC with counter-propagating signal and idler fields~\cite{christ2009pure,liu2021observation,luo2020counter,mataji2023narrow} can restrict the emission to a (nearly) single spectral mode. Similarly, domain-engineered crystals can create photons with largely reduced spectral~\cite{branczyk2010engineered,ben2013spectral,dosseva2016shaping} and spatial~\cite{baghdasaryan2023enhancing} correlations.

On the other hand, the SPDC process can be modified by embedding the nonlinear material within an optical cavity~\cite{ou1999cavity}, restricting the emission of single photons into modes with well-defined spatial and spectral characteristics. By using a tailored cavity it is possible to generate high-purity photons, avoiding filtering and therefore preserving heralding efficiency and brightness. Moreover, the spectral brightness can be increased by orders of magnitude with respect to the non-resonant case, and the generated photons exhibit reduced bandwidths, facilitating efficient interfacing with atomic or solid-state systems and reducing the effects of chromatic dispersion and path-length differences, e.g., in optical fibers. Most cavity-enhanced SPDC sources rely on an external cavity formed by mirrors~\cite{ou1999cavity, oberparleiter2000cavity, wang2004polarization, kuklewicz2006time, scholz2007narrow, bao2008generation, wang2008observation, wolfgramm2008bright, scholz2009statistics, nielsen2009time, hockel2011direct, wolfgramm2011atom, fekete2013ultranarrow, slattery2015narrow, rambach2016sub, rielander2016cavity, ahlrichs2016bright, tsai2018ultrabright, niizeki2018ultrabright,moqanaki2019novel,slattery2019background, muller2020general} (see Ref.~\cite{slattery2019background} for a review); however, SPDC embedded in an external cavity leads to bulky setups that require active cavity stabilization techniques and emit in clusters of adjacent spectral modes because they have several longitudinal modes per cluster, requiring more filtering.

In contrast, a monolithic cavity design, i.e., the crystal itself forms the cavity, with dielectric coatings at its end facets, offers improved stability, robustness, compactness, and ease of operation. In contrast to an external cavity source, it is typically sufficient to stabilize only the temperature, which can be used to select a cavity mode. Monolithic cavity sources have been studied in the context of squeezing~\cite{breitenbach1995squeezed, yonezawa2010generation, mehmet2011squeezed, ast2013high } and as a photon-pair source using waveguides~\cite{pomarico2009waveguide,pomarico2012engineering,luo2015direct,brecht2016versatile,ikuta2019frequency} and bulk crystals~\cite{chuu2012miniature,mottola2020efficient}; likewise, microring resonators have been demonstrated as narrowband photon-pair sources~\cite{fan2023multi,chen2024ultralow,zeng2024quantum}. By harnessing the clustering effect~\cite{eckardt1991optical}, designing a cavity with carefully engineered length, reflectivities and phases, single longitudinal-mode photons have been generated in waveguides~\cite{pomarico2012engineering,luo2015direct}, although they suffer from high intra-cavity losses that limit the escape efficiency, which in turn limits the accessible heralding efficiency and source brightness. Therefore, a bulk crystal monolithic cavity could offer advantages over the waveguide version. Nevertheless, the generation of spectrally single-mode photon pairs with a monolithic cavity---and the resulting indistinguishability of the heralded photon---has not been demonstrated.

In this work, we present a heralded single-photon source based on pulsed SPDC in a monolithic non-linear crystal cavity, designed for enhanced spectral and spatial purity. Heralded single photons at a wavelength of 1540~nm and a spectral bandwidth of 168~MHz have been generated, thereby being well-suited to telecom applications, minimizing fiber losses and fiber dispersion. A maximum photon heralding efficiency of 70\% (including detection losses) is reached owing to the well-defined spatial mode of the cavity that allows efficient fiber coupling. Enabled by the cavity confinement, the source produces high-purity photons, achieving a maximum spectral purity of the main cavity mode of $(96.2\pm2.7)\%$, while keeping the multi-pair emission  below 3\%, and a corresponding HOM visibility of $(91.2\pm9.3)\%$.

The present article is organized as follows. Sec.~\ref{sec:theory} introduces the theoretical framework: the joint spectral amplitude of SPDC in a monolithic cavity and the approach to generate spectrally pure photon pairs. Sec.~\ref{sec:methods} describes our monolithic cavity design and the experimental setup, while Sec.~\ref{sec:results} presents the results. The conclusions are analyzed in Sec.~\ref{sec:discussion}.

\section{Theory}
\label{sec:theory}
In SPDC, a pump photon with higher frequency $\omega_p$ spontaneously annihilates in a nonlinear crystal to create a pair of so-called signal and idler photons, with respective frequencies $\omega_s$ and $\omega_i$, conserving energy (i.e., $\omega_p=\omega_s+\omega_i$) and momentum. Neglecting higher-order contributions, the two-photon state after the interaction can be written as~\cite{grice1997spectral}
\begin{equation}
\ket{\Psi_{s,i}(\omega_s,\omega_i)} = \ket{0_s,0_i} + C\int_0^{\infty}\int_0^{\infty} d\omega_s d\omega_i \psi(\omega_s,\omega_i) \ket{1_s,1_i}.
\end{equation}
Here, $\psi(\omega_s,\omega_i)$ is the joint spectral amplitude (JSA), which captures the full spectral structure of the two-photon state. The JSA may be modeled by $\psi(\omega_s,\omega_i)=\alpha(\omega_s,\omega_i)\phi(\omega_s, \omega_i)$, where $\alpha(\omega_s,\omega_i)$ is the spectral amplitude induced by the pump field, and $\phi(\omega_s, \omega_i)$ the phase-matching function of the crystal. For a pulsed laser system with a Gaussian frequency distribution  (central frequency $\omega_{p0}$ and standard deviation $\sigma_f$), the pump-induced spectral amplitude is given by $\alpha(\omega_s,\omega_i) = \mathrm{exp}[{-(\omega_s+\omega_i-\omega_{p0})^2/2\sigma_f^2}]$. 
The phase-matching function, defined by the crystal dispersion and length $L$, may be expressed as $\phi(\omega_s, \omega_i)=\mathrm{sinc}(\Delta{k}(\omega_s,\omega_i) L/2) \exp(i\Delta{k}(\omega_s,\omega_i) L/2)$. Here, $\Delta{k}(\omega_s,\omega_i)=k_p-k_s-k_i+2\pi/\Lambda$ is the wavevector mismatch, with $k_\mu=n_\mu\omega_\mu/c$  (with $\mu=p,s,i$) and refractive index $n_\mu$, and the last term manifesting for a quasi-phase-matched crystal with poling period $\Lambda$.

\begin{figure}[h]
\centering\includegraphics[width=13cm]{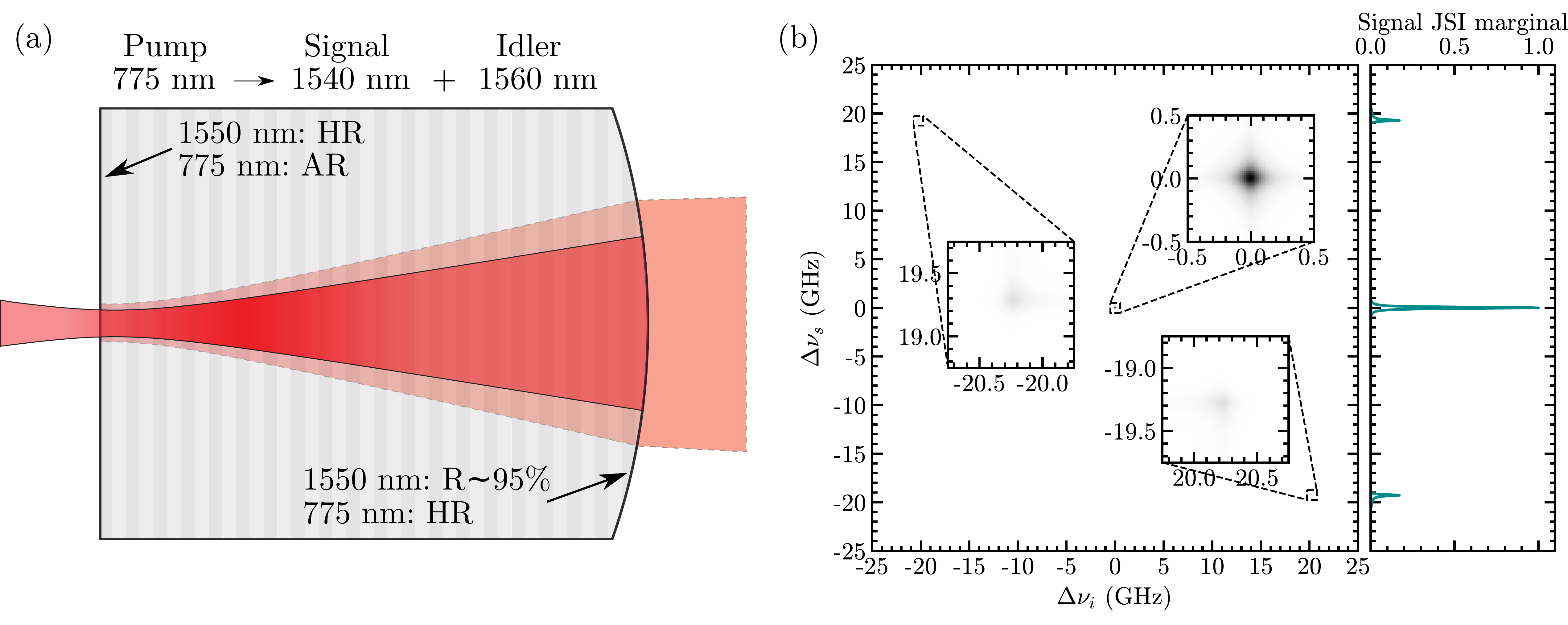}
\caption{(a) Schematic of the monolithic cavity. The nonlinear crystal with coated facets forms a cavity for the signal and idler photons. The pump photon at 775~nm is non-resonantly reflected back, while the signal and idler photons at 1540 and 1560~nm resonate within it and exit through the second facet. HR: highly-reflective coating. AR: anti-reflective coating. (b) Signal and idler cavity JSI for a 0.4~ns-long pump pulse (full width at half maximum (FWHM): 0.4~ns, 1.1~GHz) with the corresponding JSI marginal of the signal photon. The signal (idler) photons have 150~MHz (158~MHz) linewidths and 19.3~GHz (20.2~GHz) FSRs.}
\label{fig:monolithiccavityandJSI}
\end{figure}

Building a cavity around the crystal shapes the characteristics of the JSA. We consider a monolithic cavity that is resonant for the signal and idler photons with a double-pass pump, i.e., the pump is non-resonant but reflected back, as depicted in Fig.~\ref{fig:monolithiccavityandJSI}(a). The double-pass pump offers the potential for increased brightness and, if desired, narrower phase-matching bandwidth~\cite{bjorkholm2003improvement,ahlrichs2019triply}. In this configuration, the two-photon state is a superposition of photon pairs generated in the forward and backward pump directions, resulting in a modified phase-matching function that depends on the relative phase between the two pump interactions $\phi_p$ and pump reflectivity $R_p=r_p^2$. For a monolithic cavity, the JSA can be written as~\cite{jeronimo2010theory}
\begin{equation}
\psi_{\text{cavity}}(\omega_s,\omega_i)= (\mathscr{A}_s(\omega_s) \mathscr{A}_i(\omega_i))^{1/2}(1+r_p^2+2r_p\cos(\Delta k L+\phi_p))^{1/2}\psi(\omega_s,\omega_i).
\end{equation}
Here $\mathscr{A}_\mu$ is the Airy function given by $\mathscr{A}_\mu^{-1}(\omega) = 1+4\mathscr{F}_\mu^2\sin^2(\delta_\mu(\omega)/2)/\pi^2$, where $\delta_\mu(\omega)=2\theta_\mu+\delta_{\mu1}+\delta_{\mu2}$ is the phase accumulated during one pass of a field $\mu$ ($\theta_\mu$) and reflection on the mirrors ($\delta_{\mu1},\delta_{\mu2}$), and $\mathscr{F}_\mu$ the finesse of the cavity (in terms of the reflectivity R and the absorption loss $\alpha$) written as $\mathscr{F}_{\mu} = \pi(R_1R_2e^{-\alpha2L})^{1/4}/(1-\sqrt{R_1R_2e^{-\alpha2L}})$.

In conclusion, the spectrum of the cavity source is determined by the product of the pump function, the phase-matching envelope, and the frequency resonances of the cavity. Three clusters of fundamental transverse Gaussian modes (TEM00) are expected to lie within the phase-matching bandwidth of a high-finesse doubly resonant cavity for a zero net pump phase~\cite{pomarico2012engineering,luo2015direct}. The clustering effect arises from the fact that the signal and idler experience different dispersion in the crystal, which results from their orthogonal polarizations (in type-II SPDC) and small frequency non-degeneracy, and therefore have different free spectral ranges (FSR); as a result, emission occurs only in regions of the phase-matching bandwidth where the cavity resonances of signal and idler coincide. These three groups of emission peaks are commonly called clusters~\cite{eckardt1991optical}, and this effect can be exploited to generate pure longitudinal-mode single photons~\cite{pomarico2012engineering,luo2015direct}. By increasing the finesse of the cavity and maximizing the effective refractive index contrast between signal and idler modes in the monolithic design, the contributions of the immediately adjacent resonances within a cluster can be efficiently suppressed.

We theoretically analyze the two-photon state using the experimentally determined parameters of our monolithic crystal cavity (see Fig.~\ref{fig:monolithiccavityandJSI}\nobreak(a) and full description in Sec.~\ref{sec:methods}). In this analysis, we assume that the second facet is perfectly reflective for the pump ($R_p=1$), and that the pump relative phase is $\phi_p=0$, resulting in a crystal effective length of $2L$. It should be noted that in this, as in most applications, a double-pass pump is advantageous only if $\phi_p$ is zero or close to zero, since for $\phi_p=\pi$ (and similarly for $\pi/2 \leq \phi_p \leq \pi$), the two pump interactions exhibit destructive interference at $\Delta k=0$, modifying the JSA characteristics in a less favorable way. In practice, this condition is likely challenging to achieve experimentally owing to the presence of partially poled periods, but it is not detrimental to the estimation of the spectral purity within a single cluster. Furthermore, we assume that there are no phase contributions from the reflection at the facets, i.e., $\delta_{\mu1}=\delta_{\mu2}=0$ for $\mu=s,i$, and a loss per meter of $\alpha=0.1$. The refractive indices are calculated using the Sellmeier equations from Refs.~\cite{fradkin1999tunable,konig2004extended} and the temperature dependence from Ref.~\cite{emanueli2003temperature}. The corresponding signal and idler joint spectral intensity (JSI) $|\psi_{\text{cavity}}(\omega_s,\omega_i)|^2$ is shown in Fig.~\ref{fig:monolithiccavityandJSI}\nobreak(b). Most of the emission is into the central cavity mode, with a smaller portion into the two adjacent cavity modes.  In the above-considered case, the signal and idler photons have a bandwidth, determined by the cavity linewidths of 150~MHz and 158~MHz, respectively, and an FSR of 19.3~GHz and 20.2~GHz, corresponding to a cavity finesse of 128. 

\begin{figure}[h]
\centering\includegraphics[width=13cm]{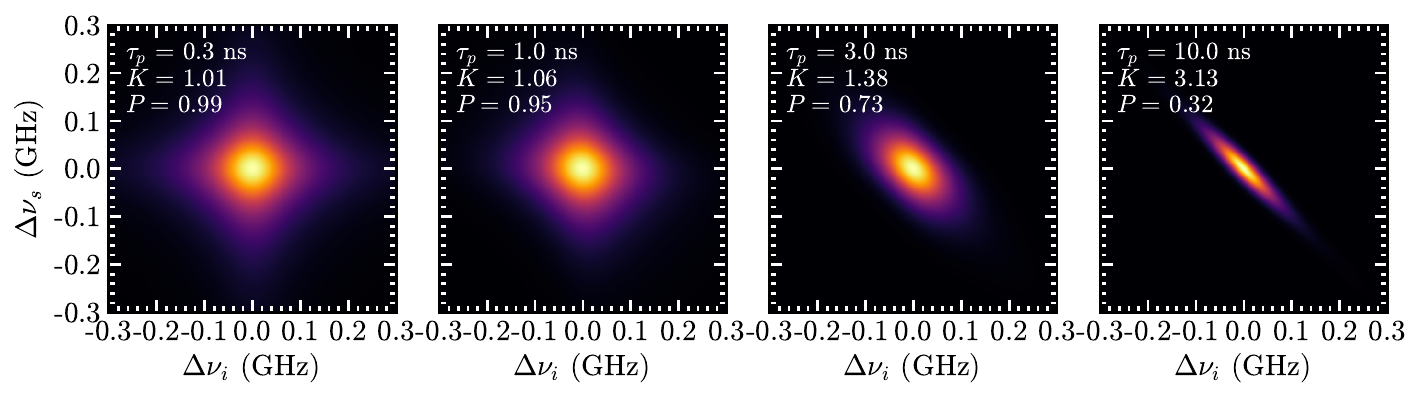}
\caption{Signal and idler cavity JSI of the central mode as a function of the frequency detuning of signal and idler for different pump pulse lengths $\tau_p$ (FWHM). The Schmidt number $K$ and spectral purity $P$ for each pump pulse length are indicated, showing that shorter pulses decrease frequency correlations.}
\label{fig:JSI_vs_pplength}
\end{figure}

To quantify the separability of the JSI for the central cavity mode, a Schmidt decomposition is performed~\cite{law2000continuous}, allowing for a theoretical evaluation of the number of frequency modes present in the state and, consequently, its spectral purity. The two-photon state is expressed in terms of orthonormal Schmidt modes $\ket{\psi_s^k(\omega_s)}$ and $\ket{\psi_i^k(\omega_i)}$ as
\begin{equation}
\ket{\Psi_{s,i}(\omega_s,\omega_i)} = \sum_k \sqrt{\lambda_k} \ket{\psi_s^k(\omega_s)}\ket{\psi_i^k(\omega_i)},
\end{equation}
where the Schmidt number $K=1/\sum_k \lambda_k^2$ is the inverse of the spectral purity P ($P=1/K$) and indicates the number of active Schmidt frequency modes, being $K=1$ ($P=1$) for a state without spectral correlations, i.e., a spectrally pure state. Figure~\ref{fig:JSI_vs_pplength} presents the signal and idler JSI for the central cavity mode for various pump pulse durations and the corresponding Schmidt number and spectral purity. For long pump pulses, the residual correlations in the JSI stem from the fact that the uncertainty in the pump energy lies below that of the signal/idler bandwidth. They can be minimized by utilizing pump pulses spectrally broader than the cavity mode, so that the JSI of the photon pair is dominated by the single-mode of the cavity rather than the pump spectral amplitude (at the expense of reduced brightness, as the cavity then excludes a fraction of the pump). Optimizing the pump pulse allows for a separable JSI (uncorrelated idler and signal frequencies), making possible the generation of spectrally pure heralded single photons. A spectral purity of $99.4\%$ (Schmidt number $K=1.01$) can be reached using a 0.3~ns-long pump pulse. It must be noted that, while the central mode is pure, correlations are introduced by distributing some amplitude into the neighboring spectral modes; however, these can be spectrally filtered out. Without filtering, i.e., with the contribution of the adjacent cavity modes, the maximum achievable spectral purity is 79.4\% ($K=1.26$) using a 1.1~ns pump pulse; although the output is not single-mode, its purity is significantly higher than that of a non-engineered SPDC source. It is important to emphasize that the filtering here is only used to select one of the cavity longitudinal modes within the phase-matching bandwidth for single-mode operation rather than to make the bandwidth arbitrarily narrow; unlike other filtered single-photon sources, the selected photon cavity modes are not spectrally truncated. This indicates that cavity-enhanced SPDC with pulsed pumping can in principle result in arbitrarily high spectral purity when sufficiently short pump pulses are used and a single longitudinal mode is spectrally selected.

\section{Methods}
\label{sec:methods}
The here presented photon source is formed from a periodically-poled potassium titanyl phosphate (ppKTP) crystal with a poling period of 45.35~µm, a length of 4.2~mm, and face dimensions of 1x2~mm$^2$. This crystal is phase-matched to convert pump photons at 775~nm into pairs of signal and idler photons at 1540~nm and 1560~nm, respectively, via type-II down-conversion. The facets of the crystal are cut, polished, and coated to form a plano-convex cavity with a 10~mm radius of curvature for the signal and idler photons, as illustrated in Fig.~\ref{fig:monolithiccavityandJSI}\nobreak(a).  The planar facet is coated to achieve a reflectivity of 99.9\% (0.2\%) for the signal/idler (pump) photons, while the convex facet served as the output coupler, coated to achieve a reflectivity of 95.4\% (99.8\%) for the signal/idler (pump) photons. The crystal length, out-coupler reflectivity, and radius of curvature are chosen with consideration of the escape efficiency, the linewidth of the photons, the focal parameter, and the cavity stability.

\begin{figure}[h]
\centering\includegraphics[width=12cm]{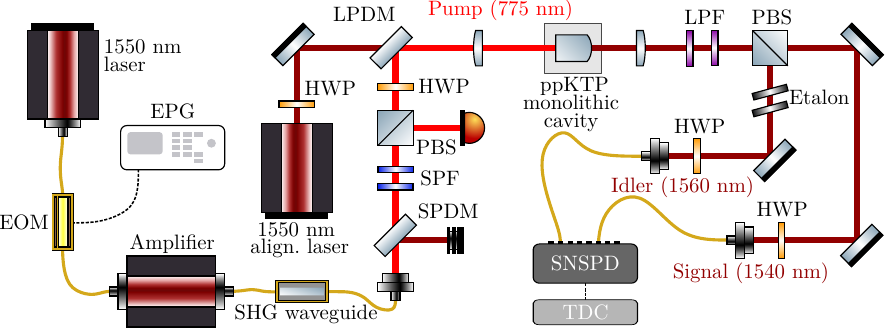}
\caption{Experimental setup. The setup consists of the generation of pump pulses at 775~nm via second harmonic generation and of the generation of signal and idler photon pairs at 1540~nm and 1560~nm, respectively, via SPDC in the monolithic crystal cavity shown in Fig.~\ref{fig:monolithiccavityandJSI}\nobreak(a). EOM: electro-optic modulator, EPG: electrical pulse generator, SHG: second harmonic generation, SPDM: shortpass dichroic mirror, LPDM: longpass dichroic mirror, SPF: shortpass filter, LPF: longpass filter, PBS: polarizing beamsplitter, HWP: half-wave plate, SNSPD: superconducting nanowire single-photon detector, TDC: time-to-digital converter.}
\label{fig:experimentalsetup}
\end{figure}

Figure~\ref{fig:experimentalsetup} shows the setup of the experiment. The generation of pump light at 775~nm is done via second harmonic generation (SHG)~\cite{franken1961generation,kleinman1962theory} from 1550~nm light. Nanosecond pulses are carved from a 1550~nm continuous-wave (CW) distributed-feedback erbium fiber laser using a fiber-coupled electro-optic modulator (EOM), and then amplified via an erbium fiber amplifier. The amplified 1550~nm pulses are subsequently converted to 775~nm by a fiber-coupled SHG waveguide. The waveguide is made of 5\% magnesium-doped poled lithium niobate, with a length of 30~mm and a poling period of 18.5~µm. The transmission of non-converted 1550~nm light is suppressed in free space using a short-pass dichroic mirror followed by two short-pass filters. This setup can generate optical pulses at 775~nm with durations ranging from 0.1 to 10~ns, repetition rates of up to 1~GHz and average powers in the tens of mW. The imperfect extinction of the EOM is partially mitigated by the quadratic dependence of SHG on the input power~\cite{bloembergen1962light}.

The pump light is collimated from a single-mode fiber with an aspheric lens of 12~mm focal length and overlapped with the mode of the cavity with a plano-convex lens of 125~mm focal length, resulting in an expected beam waist of 26~µm at the planar facet of the crystal. The crystal is positioned in a temperature-controlled mount at $46.54^{\circ}\text{C}$, maintaining temperature stability to a precision of 1~mK. The down-converted light corresponds to the TEM00 Hermite-Gaussian modes of the optical cavity for horizontal and vertical polarization, with predicted signal and idler beam waists of 37~µm and 38~µm at the planar crystal facet, respectively, which are then collimated with a plano-convex spherical lens of 75~mm focal length and filtered with two longpass filters to block the residual transmission of the pump. The signal and idler photons are spatially separated with a polarizing beam splitter, and the idler photon is filtered with two angle-tuned etalons to remove adjacent spectral modes and clusters (measured transmission bandwidths of 5 and 14~GHz respectively). Both fields are focused with an aspheric lens of 11~mm focal length and coupled into anti-reflection-coated single-mode fibers with over 90\% efficiency using a three-axes flexure stage that nominally provides 20~nm adjustment resolution. They are detected with superconducting nanowire single-photon detectors.

\section{Results}
\label{sec:results}
Using a CW pump we achieved heralding efficiencies of 70$\%$ in the idler arm and 15$\%$ in the signal arm, respectively, including transmission and detection losses. Here, the heralding efficiency is the probability that the partner photon is detected in a fiber once its counterpart has been detected. The asymmetry arises from filtering applied only to the idler arm, which intersects the central cavity mode that couples efficiently into single-mode fibers, thereby increasing the idler heralding efficiency while leading to a lower signal heralding efficiency. From the obtained efficiencies, we estimate the cavity escape efficiency to be about 94\%. The detected signal-idler pair rate was $8.1\times10^3~\mathrm{pairs\,s^{-1}\,mW^{-1}}$, corresponding to a photon generation rate inside the cavity of $7.4\times10^4~\mathrm{pairs\,s^{-1}\,mW^{-1}}$. With a pulsed pump, when the pulse bandwidth is larger than the cavity linewidth, the cavity (and filter) excludes a fraction of the pump, leading to a brightness and heralding efficiency reduction that inversely scale with the pump pulse length. For instance, the brightness reduction is only about 20\% for pump pulses of 5~ns length, increasing to roughly 50\% for 1.5~ns pulses or 75\% for 0.5~ns. The idler heralding efficiency remains approximately independent of the pump length while the signal heralding efficiency follows a behavior similar to that of the brightness. Furthermore, the brightness remained stable for several hours without any adjustments or tuning.

\begin{figure}[h]
\centering\includegraphics[width=11.5cm]{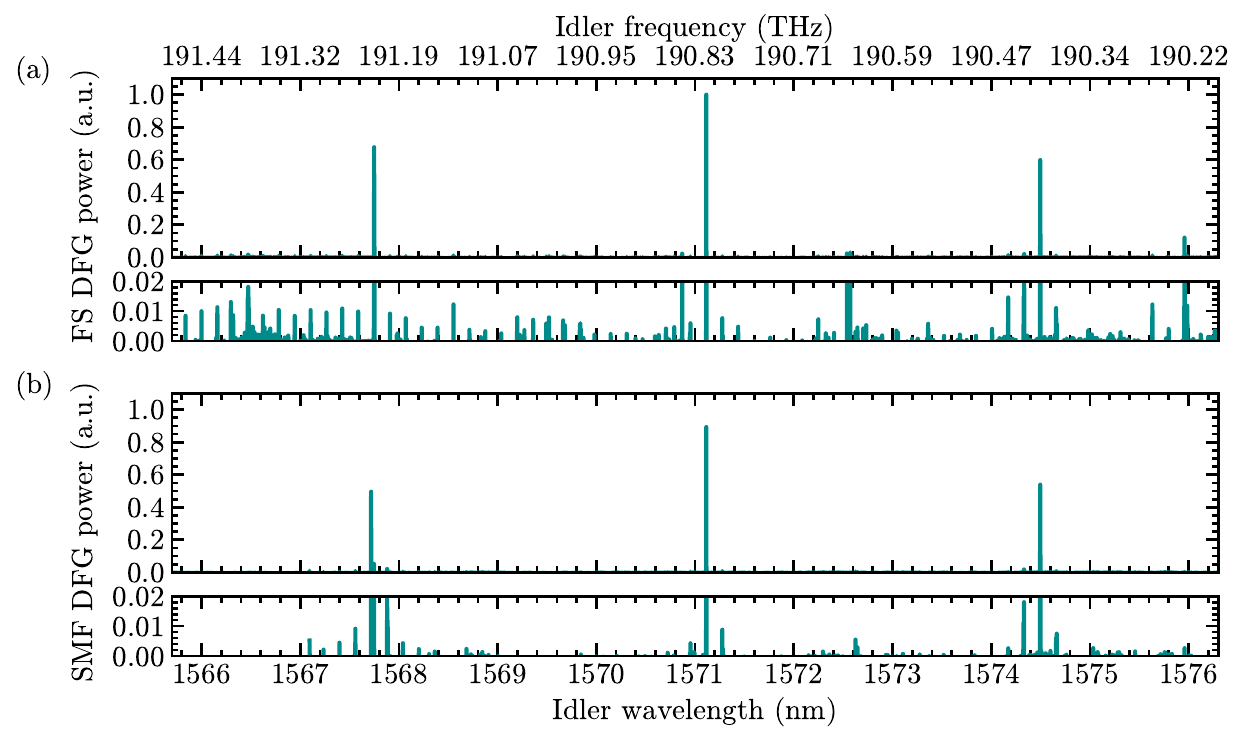}
\hspace{0.7cm}
\caption{Idler spectrum measured via difference frequency generation. Measured idler power for (a) free-space detection and (b) spatially filtered detection (after a single-mode fiber). For a fixed pump wavelength, the signal mode was seeded with a wavelength-tunable CW laser and scanned across the phase-matching bandwidth. The full-scale plot shows three clear clusters for both detection methods. However, a small but consistent background with contributions from higher-order modes (zoomed-in plots) is clearly evident in the free-space detection (a) and is dramatically reduced with single-mode fiber detection (b).}
\label{fig:DFGspectrum}
\end{figure}

To gain insights into the spectral properties of the source, the idler spectrum was directly measured via difference frequency generation (DFG). The pump beam, generated with an external-cavity diode CW laser at 780.25~nm, and a tunable CW seed beam, tunable over a $\pm$6~nm range around 1550~nm, were overlapped in free space through a dichroic mirror and injected into the crystal, generating the idler field. The residual seed light, which was transmitted due to the imperfect extinction of the polarizing beam splitter, was filtered with an additional bandpass filter. The transmission of the cavity at the idler arm, measured before and after a single-mode fiber, is shown in Fig.~\ref{fig:DFGspectrum}. The spectrum exhibits three TEM00 modes---the three primary peaks that exclusively couple to single-mode fibers---consistent with the mode clustering effect in a doubly-resonant cavity of high finesse (see Sec.~\ref{sec:theory}). Nevertheless, the intensity of the side frequency clusters is not as low as desired due to the double-pass pump accumulating a phase, likely because of a residual crystal length (i.e., an intermediate length of crystal after the last full poling), effectively broadening the phase-matching bandwidth, leaving three dominant TEM00 spectral modes. The main frequency TEM00 mode was coupled into single-mode fiber with above 90\% efficiency. The additional little peaks in the background (Fig.~\ref{fig:DFGspectrum}\nobreak(a)\nobreak(bottom)) represent the weakly phase-matched emission of the idler light into higher-order modes, although the seed (signal) and pump light are both largely in the fundamental mode. We believe this small but largely constant background of high-order modes, essentially stemming from a phase-matching process where only one of the signal or idler photons will ever couple efficiently into single-mode fiber, explains the observed limitations in the heralding efficiency~\cite{bennink2010optimal}.

\begin{figure}[h]
\centering\includegraphics[width=13.2cm]{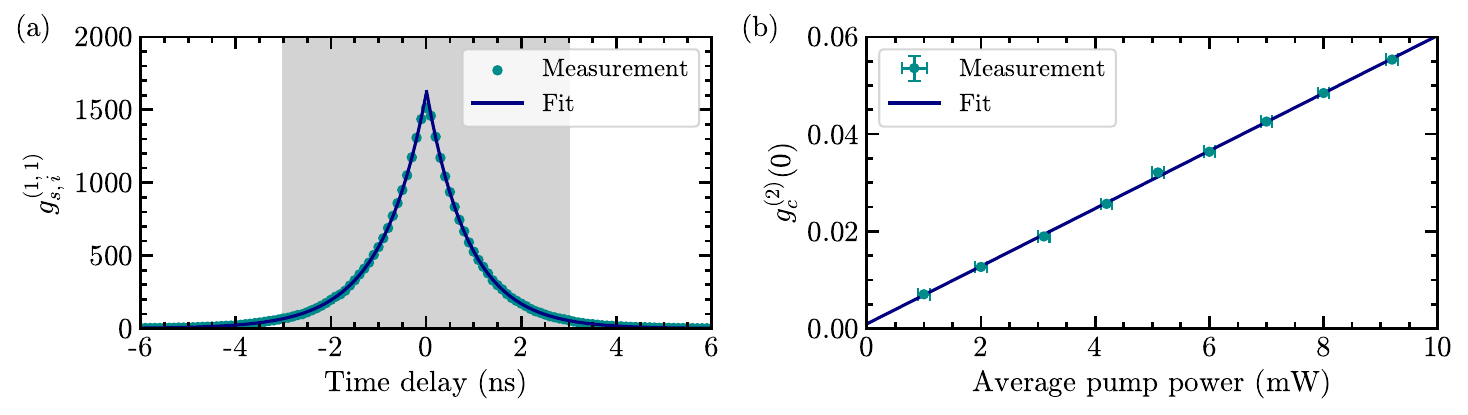}
\caption{(a) Measured signal and idler cross-correlation over their temporal separation for 2.03~mW pump power. The exponential temporal function of each photon is fit to the data. The integration window of 6~ns for all subsequent measurements is shown in gray. (b) Measured heralded-signal autocorrelation as a function of the average pump power with the corresponding linear fit. The error bars are derived from Poissonian photon counting statistics.}
\label{fig:correlfunctions}
\end{figure}

Using second-order correlation measurements, we characterized the statistical properties of the photons generated by the source. Firstly, we measured the second-order cross-correlation $g_{si}^{(1,1)}$ between signal and idler photons using a CW pump (power of 2.03~mW), based on the timing of their detection, as shown in Fig.~\ref{fig:correlfunctions}\nobreak(a). The prominent peak at zero time delay reflects the emission of photons in pairs, with a temporal profile arising from the correlation of the exponential temporal structure of both photons mirroring the Lorentzian cavity spectral modes. We fit to the data the rising and falling exponential of each photon~\cite{scholz2009statistics} $\exp{[2\pi\Delta\nu_s t] }\Theta(-t)+\exp{[-2\pi\Delta\nu_i t]}\Theta(t)$ in terms of the Heaviside function $\Theta$, from which we infer photon bandwidths for the signal and idler of $\Delta\nu_s=167.9\pm0.2$~MHz and $\Delta\nu_i=180.4\pm0.2$~MHz (respective coherence times of 1.31~ns and 1.22~ns). Furthermore, in order to verify the single-photon nature of the heralded signal photons, we measured the signal second-order autocorrelation conditioned on the detection of an idler photon $g_c^{(2)}$ using a Hanbury Brown and Twiss (HBT) fiber interferometer in the signal arm, and observed that the multi-photon generation for 0.96~ns pump pulses at a 20~MHz repetition rate remained below 3\% for average pulsed pump powers up to 5~mW, see Fig.~\ref{fig:correlfunctions}\nobreak(b). 

\begin{figure}[h]
\centering\includegraphics[width=13.2cm]{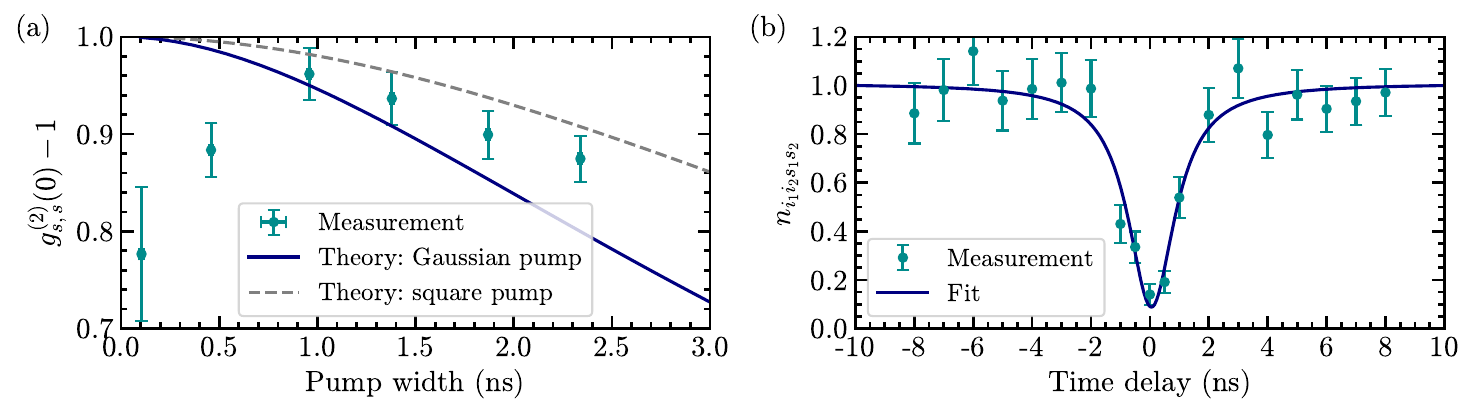}
\caption{(a) Measured spectral purity of the central cavity mode as a function of the pump pulse length (FWHM). The theoretical models for Gaussian and square pump pulses are illustrated. (b) Measured four-photon event rate (normalized) using a Hong-Ou-Mandel interferometer, and scanned over the temporal separation of the two photon pairs. Pump pulses of 0.96~ns duration at an average power of 2~mW were utilized. The integration time for each data point is 10~min. An inverse Lorentzian function is fitted to the data points. The error bars in (a) and (b) are derived from Poissonian photon counting statistics.}
\label{fig:spectralpurity}
\end{figure}

We subsequently measured the unconditioned second-order autocorrelation $g_{ss}^{(2)}$ of the signal using an HBT fiber interferometer in the signal arm. This measurement provides an indirect way to estimate the effective number of modes $K$ as $g_{ss}^{(2)}(0)\approx1+1/K$~\cite{christ2011probing}. Figure~\ref{fig:spectralpurity}\nobreak(a) presents the measured spectral purity of the central cavity mode for different pump pulse lengths and the corresponding simulated theoretical curves for Gaussian and square pump pulses, since, due to the pulse shaping characteristics of our pump system, the longer pulses have a temporal profile that lies between a Gaussian and a square pulse. The measured values align with the expected spectral decorrelation by pump pulses broader than the cavity mode discussed in Sec.~\ref{sec:theory}, reaching an experimental maximum spectral purity of $P_\mathrm{M} = (96.2\pm2.7)\%$ using a 0.96~ns pump pulse. For the shortest pump pulses, the purity decreases because the broader pulses have some non-zero overlap with the adjacent modes, and the two etalons are insufficient to filter them out; moreover, when the pulses get shorter, the noise floor becomes more significant. Consequently, spectral purities approaching 100\%, as predicted by theory, cannot be realized.

Finally, as a direct measure of the total state purity, we quantified the indistinguishability of the generated single photons through the HOM effect~\cite{hong1987measurement}. We investigated the interference between two heralded single photons produced by consecutive pump pulses by varying their temporal separation utilizing an unbalanced Mach-Zehnder fiber interferometer in the signal arm. The polarization in the interferometer was adjusted using fiber polarization controllers and classical light, achieving an interference contrast of $(96.4\pm0.2)\%$, constrained by the polarization state and imperfections in the beam splitting ratio. We employed 0.96~ns-long pump pulses---offering the highest spectral purity---at an average power of 2~mW, corresponding to a single-photon Fock state purity of $(98.7\pm0.1)\%$. We varied the repetition rate of the pump around 20.33~MHz to adjust the temporal separation of the photons. The normalized rate of four-fold events (the two signals and the corresponding two heralding idlers) at the two outputs of the interferometer, as a function of the pump pulses separation, can be seen in Fig.~\ref{fig:spectralpurity}\nobreak(b). In this measurement, no additional temporal filtering was applied, and a coincidence window of 6~ns was used, as displayed in Fig.~\ref{fig:correlfunctions}\nobreak(a). We see a pronounced bunching at zero time delay, demonstrating the high indistinguishability of the heralded photons from our monolithic cavity. By fitting an inverse Lorentzian to the data, we obtain a HOM visibility of $V_\mathrm{HOM}=(91.2\pm9.3)\%$. The observed HOM visibility is consistent with the product of the spectral purity, photon-number purity, and interferometer contrast. The large statistical error arises from the low probability of generating two consecutive photon pairs and successfully detecting all four photons in a passive interferometer. Furthermore, it should be emphasized that $V_\mathrm{HOM}$ is partially limited by the experimental interferometer; an improved version of the interferometer, with nearly 100\% contrast, is expected to yield HOM visibilities of up to 95\%. Similarly, $V_\mathrm{HOM}$ is also degraded by multi-pair SPDC emissions. Both contributions set a lower bound for the photon indistinguishability $V_\mathrm{HOM}$, while the upper bound is determined by the maximum achieved spectral purity $P_\mathrm{M}$.

\section{Discussion}
\label{sec:discussion}
Here, we have presented a heralded, pulsed SPDC single-photon source based upon a monolithic cavity geometry. The monolithic cavity generates narrowband photons (168~MHz spectral bandwidth) at the telecom C-band (1540~nm wavelength), allowing efficient transmission of these photons in optical fibers with minimal fiber chromatic dispersion. The well-defined spatial modes of the cavity enable efficient fiber coupling, achieving a maximum heralding efficiency of 70\% including losses in detection. The multi-photon contamination remained below 3\% for pump powers up to 5~mW, as expected from SPDC photon-number statistics. The cavity generates photons with an estimated  $(96.2\pm2.7)\%$ spectral purity, and the indistinguishability was further quantified through the HOM effect, achieving a visibility of $(91.2\pm9.3)\%$. 

In upcoming experiments we will design a cavity with both facets non-reflective for the pump. Here, the pump accumulates a phase when it reflects that broadens the phase-matching bandwidth, increasing the intensity of the side cavity clusters. Since the control over this phase in the manufacturing and coating of the crystal can be challenging, we believe that a single-pass pump offers a more straightforward approach. Additionally, a higher cavity finesse could suppress the adjacent cavity modes within a cluster and generate unfiltered single-mode photons, at the cost of worse heralding efficiency.

Overall, the potential of this source to produce high-purity photons with high efficiency makes it a promising candidate for scalable and practical quantum information processing applications.

\begin{backmatter}
\bmsection{Funding}
Bundesministerium für Forschung, Technologie und Raumfahrt (13N15870).

\bmsection{Acknowledgment}
The authors thank Leon Messner (AQLS) for the design of the crystal oven. 

\bmsection{Disclosures}
The authors declare no conflicts of interest.

\bmsection{Data Availability Statement}
Data underlying the results presented in this paper are not publicly available at this time but may be obtained from the authors upon reasonable request.

\end{backmatter}

\bibliography{references}

\end{document}